\documentclass[12pt]{article}
\usepackage{amsmath,amssymb,epsf,epsfig,graphics,graphicx}

\def \be {\begin{equation}}
\def \ee {\end{equation}}
\def \bea {\begin{eqnarray}}
\def \eea {\end{eqnarray}}

\def \dels {\partial\kern-.5em / \kern.5em}
\def \As {{A\kern-.5em / \kern.5em}}
\def \Ds {D\kern-.7em / \kern.5em}

\begin{document}
\begin{titlepage}
\bigskip
\bigskip\bigskip\bigskip\bigskip
\bigskip \bigskip
\centerline{\Large \bf {CMB Constraints on the Holographic Dark Energy Model}}
\bigskip
 
\bigskip\bigskip

\centerline{ \large Hsien-Chung Kao\footnote{hckao@phy.ntnu.edu.tw},  Wo-Lung Lee\footnote{leewl@phy.ntnu.edu.tw}, Feng-Li Lin\footnote{linfengli@phy.ntnu.edu.tw}}
\bigskip
\centerline{\large \it Department of Physics, National Taiwan Normal University}
\centerline{\large \it Taipei, Taiwan 116}

\bigskip\bigskip

\bigskip\bigskip

%ABSTRACT

\begin{abstract}
\medskip
\noindent We calculate 
%LIN
the angular scale of the acoustic oscillation from the BOOMERANG and WMAP data on the cosmic microwave background (CMB) to constrain the holographic dark energy model recently proposed by Li. 
We find that only the phantom-like holographic dark energy survives the cosmological tests. 
%lin
This is, however, inconsistent with the positive energy condition implicitly assumed in constructing Li's model. Therefore the model is marginally ruled out by the present CMB data. As a supplementary check, we also calculate the suppression of the matter density fluctuation due to the late time integrated Sach-Wolfe effect by the holographic dark energy, the result is within the tolerance of the cosmic variance.  Some aspect about the saturation of the cosmic holographic bound is also discussed.\end{abstract}
\end{titlepage}

\setcounter{footnote}{0}

%%%%%%%%%%%%%%%%%%%%%%%%%%%%%%%%%%%%%%%%%%%%%%%%%%%%%%%%%%%%%%%%%%%%%%
%%%%%%

\section{Introduction and Conclusion}

  The idea of holographic principle for strong gravity systems inspired from the area law of Bekenstein-Hawking's entropy has been widely accepted after the success of the AdS/CFT correspondence. The extension of the holographic principle to more general cases such as the FRW background was first proposed by Fischler and Susskind \cite{fs}, clarified by Bak and Rey \cite{br} and completed later on by Bousso \cite{bousso} in the name of covariant entropy bound. It states that the entropy on the non-expanding light-sheet of a surface $B$ will be bounded by the area of $B$. The intuitive picture behind the statement is that the more entropy we put in a region, the more we increase the energy in it \footnote{Here, one must assume the matter associated with it obeys the positive energy condition. In contrast, the phantom-like dark matter/energy violates the positive energy condition and does not have to obey the holographic bound.}. This in turn will curve the passing-by light ray more and eventually a caustic will form and terminate the light-sheet. The maximal entropy one can have is for the whole region to collapse into a black hole, and thus the entropy is bounded by Bekenstein-Hawking's area law.
  
  It is then intriguing if the generic covariant entropy bound can be used to constrain cosmological models. Recently, a viable model for the dark energy based on the holographic principle has been proposed by Li \cite{li1} (see also the follow-up in \cite{li2}.) The model seems to comply with the recent observation that the Universe is accelerating \cite{nova}.  Therefore, it deserves further check to see whether its prediction is also consistent with the recent data \cite{boom,wmap} on the cosmic microwave background (CMB) anisotropies. As far as we can see, the recent CMB data has become precise enough to constrain the model to some extent.
  
  In this note, we check the so called HCDM (holographic dark energy plus cold dark matter) model by first calculating the angular scale of the acoustic oscillation on the last scattering surface, and then estimating the integrated Sach-Wolfe (ISW) effect for the growth of the matter density contrast. We find that the recent cosmological data prefer phantom-like holographic dark energy. Similar conclusion was also arrived in the first reference in \cite{li2} by examining the luminosity distance of the SN Ia data \cite{nova}. Unfortunately, the phantom-like dark energy is inconsistent with the positive energy condition (PEC), which must be satisfied to derive Li's holographic bound.  We must then conclude that the HCDM is disfavored by the CMB data from our analysis. Moreover, the phantom-like HCDM will also violate the generalized 2nd law of thermodynamics (GSL) if we assume that the later universe is dominated by the holographic dark energy whose entropy is bounded by the area law.

%lin   
%  HCDM model will generally At first glance, it seems possible to circumvent this difficulty even though the bound decreases with time since the entropy of the holographic dark energy may increase without exceeding the area law bound. A closer look, however, shows that even if the entropy of the holographic dark energy dose not saturate the usual area law bound, it will still be proportional to three-quarter power of the area at least \cite{ckn}. Hence, the GSL will eventually be violated for the phantom-like HCDM.
  
%lin  
%Moreover, we will also point out that the positive energy condition (PEC) for the vacuum energy is assumed in deriving the holographic bound, therefore the phantom-like HCDM violating the PEC is an inconsistent model.

%kao 
%One possible resolution to the difficulty is that future CMB and SN Ia data are significantly improved and overturn the conclusions we arrive at in this note. Even if this is not the case, it does not necessarily mean holographic bound is ruled out. Instead, it only implies that we need to be more careful in imposing the holographic bound. 
    
In the literature, when the holographic principle is exploited to build HCDM models, there is always an implicit assumption, which we call ``maximal darkness conjecture" (MDC), that the holographic bound is saturated so that there is a definite UV-IR connection. However, there is no strong argument that our universe is right on the verge of saturating the holographic bound.  The area law conjectured in \cite{fs,bousso} is used in the naive version of MDC. It leads to the conclusion that the nature is filled with the maximal amount of dark matter or dark energy so that our universe has become a black hole. In this form, the conjecture is hardly justifiable and does not yield a realistic model as will be discussed in the next section. To have a more realistic HCDM model, Li uses the area law suggested in \cite{ckn}.  This leads to a new bound related to Jean's instability. Although the new bound is less restrictive, it still implies that our Universe is on the verge of becoming a black hole. This is again hard to justify. Despite of the problem, the resultant model seems to be viable.
   
   From the above discussion it is clear that there may still be space to further relax the MDC so that a more realistic non-phantom like dark energy model based on holographic principle could emerge.  A most intuitive evidence that the MDC could be wrong is that there is no single evidence that we are approaching a black hole singularity anytime soon. However, we will leave the justification/falsification for MDC to future works and focus on the cosmological test of the HCDM models. 
   
   In the next section we will review the HCDM model and give some details to make the above discussions more concrete. Especially the requirement of PEC in constructing HCDM is noted. In the last section we will provide our analysis of the CMB constraints on the HCDM model, we find that the CMB data 
rule out HCDM since our result favors the phantom-like dark energy which violates PEC.
In addition, we also calculate the matter density suppression due to the late time ISW effect. We find that the suppression is within the tolerance of cosmic variance, which is about $30\%$ difference in amplitude of the matter density in $\Lambda$CDM. Since the cosmic variance is large, we will use the matter suppression not as a crucial but as a consistent test for HCDMs. The result of the test would help us visualize how the running of dark energy will affect the CMB fluctuation.

\section{The model}
  
      The basic idea of the holographic dark energy is that we can use the saturation of the entropy bound to relate the unknown UV scale $\Lambda$ to some known cosmological scale, so that we can have a viable formula for the dark energy which is of quantum gravity origin characterized by $\Lambda$. Naively one may choose the covariant entropy bound to get the UV-IR connection, however, this would imply that the whole universe is dominated by black hole states. In other words, the back-reaction caused by the quantum dark energy will be too large for classical gravity to be reliable.

%kao
   The authors in \cite{ckn} proposed a more restrictive bound 
\be\label{bound1}
L \gtrsim G_N E_{\Lambda} \;\; \Longrightarrow \;\; L\Lambda^2 \lesssim M_p,
\ee
where $L$ is the size of the system, $E_{\Lambda}=L^3 \Lambda^4$ and $8 \pi G_N=1/M_p^2$. The above argument bases on the bending of light-rays due to the gravitational energy $\rho+p$. One must assume that the vacuum energy does obey the positive energy condition (PEC), i.e., $\rho_{\Lambda}+p_{\Lambda}>0$. Otherwise, the vacuum energy will unbind the light-rays, and contradicts the statement that more entropy implies more energy.\footnote{This can be seen by examining the Raychaudhuri equation. The same PEC is imposed when deriving the covariant entropy bound \cite{br,FMW}.} This then rules out the cosmological constant and the phantom-like dark energy.

%kao%lin
Equation (\ref{bound1}) means that no gravitational collapse will be induced by the vacuum energy so that we can still trust classical gravity. From UV-IR connection, the saturation of the bound (\ref{bound1}) implies that the vacuum energy density is given by
\be 
%kao
\label{rhoL}
\rho_{\Lambda}=3c^2 M_p^2 L^{-2}. 
%kao
\ee
%kao
Here, the parameter $c$ characterizes the degree that the Jeans bound is saturated.  
%lin
In the above we do not distinguish the energy density $\rho$ from the gravitational energy density $p+\rho$ in deriving (\ref{rhoL}) from (\ref{bound1}). The difference is not essential as long as the equation of state $w_{\Lambda}=p_{\Lambda}/\rho_{\Lambda}$ is of order one such that $\rho_{\Lambda}$ and $\rho_{\Lambda}+p_{\Lambda}$ are of the same  order of magnitude. This is indeed the case for HCDM as we shall see in equation (\ref{eos}) that $w^{0}_{\Lambda}<w<-1/3$ where $w^{0}_{\Lambda}$ is today's value of the equation of state for dark energy.

As discussed in the last section, there is an implicit assumption in the ``maximal darkness conjecture" (MDC) where the proposed bound is saturated so that one can derive the UV-IR connection. The MDC is at work if $c$ is of order one, and this turns out to be the case from data constraints.

%kao%lin
Although the form of equation (\ref{rhoL}) seems quite universal and may apply to any system such as the galaxy clusters besides the universe, equation (\ref{bound1}) suggests that for a fixed UV-cutoff $\Lambda$ the bound is harder to be saturated in smaller systems. Therefore we will assume it is saturated only for the cosmic scale with a UV-cutoff $\Lambda$ of quantum gravity origin.

%kao

   We should choose an appropriate IR scale $L$ to make the model compatible with the observed dark energy. If $L$ is the Hubble horizon, namely $L\sim 1/H$, then this result is quite suggestive since it agrees with what we observe today, i.e. the Universe is dominated by the dark energy so that $\rho_{\Lambda} \approx \rho_c\equiv 3M_p^2 H^2$. Moreover, the entropy of the dark energy \cite{ckn} is 
\be\label{bound2}
S_J=(S_{CEB})^{3\over 4}
\ee
which is below Bousso's covariant entropy bound $S_{CEB}$. In contrast, saturating Bousso's bound will yield $\rho_{\Lambda}\approx (M_p^2H)^{4\over 3}$, which is inconsistent with the observational data today unless some extreme fine-tuning is done. 

  We know that today's Universe is accelerating according to SN Ia data \cite{nova}. It then requires that $w_{\Lambda}\equiv p_{\Lambda}/\rho_{\Lambda} <-1/3$. It is easy to see that the naive choice of $L\sim 1/H$ will give $w_{\Lambda}=0$ \cite{hsu}. To have a model  consistent with the previous condition, Li \cite{li1} proposed that one should adopt the covariant entropy bound and choose $L$ to be either the particle horizon $R_H(t)=a(t)\int^t_0 {dt' \over a(t')}$ or the event horizon $R_h(t)=a(t)\int_t^{\infty} {dt' \over a(t')}$. Making use of energy conservation, we then obtain the changing rate of the horizon size   
\be
{d R_{H,h} \over d a}={R_{H,h} \over a}\left(1\pm \sqrt{\Omega_{\Lambda}\over c^2}\right),
\ee   
and the equation of state
\be\label{eos}
w_{\Lambda}=-{1\over 3}\left( 1\mp 2 \sqrt{\Omega_{\Lambda}\over c^2}\right)\;.
\ee
Here $a$ is the scale factor of the Robertson-Walker metric, and the upper(lower) sign is for $L=R_H(R_h)$. It is then obvious that using $R_H$ as $L$ will not lead to an accelerating Universe. Moreover, for $L=R_h$ we see that $w^{0}_{\Lambda} < w_{\Lambda}<-1/3$, the upper bound is attained when $\Omega_{\Lambda}$ approaching zero in the early Universe. Therefore, the magnitude of $w_{\Lambda}$ is always of order one from the early Universe till today.

%kao%lin
  Although using $R_h$ as the comic scale can give rise to an accelerating Universe, it is phantom-like today, i.e., $w^{0}_{\Lambda} <-1$ if $c^2 < \Omega^0_{\Lambda}\approx 0.7$. In such case, both the GSL and PEC will be violated in this model. The violation of GSL can be seen explicitly by calculating the cosmic entropy associated with the dark energy is $S_J=(\pi R^2_h)^{3\over4}$ as given in (\ref{bound2}). Since $\Omega_{\Lambda}$ approaches 1 in the later Universe, we must require $c^2 \ge 1$ so that $dR_h/ da\ge 0$ all the time and the GSL and PEC are always obeyed. We need the PEC to derive the bound of energy density (\ref{rhoL}), which is the basis of the HCDM model. Therefore, it does not make sense to consider phantom-like HCDM model.

%lin  
%Of course, such conclusion can be achieved only if we assume that there is no %unexpected source of additional entropy from the novel physics. 
%kao

\section{The constraints from CMB data}

  Now we would like to use the CMB data to further constrain the HCDM model parameter $c$. Instead of doing the global fitting to the CMB data, we will choose two relevant parameters synthesizing the essence of CMB physics, namely, the acoustic scale characterizing the acoustic oscillations in the recombination epoch, and the growth of the matter density under the influence of the ISW effect.
  
  The Friedmann equation in the flat Robertson-Walker metric is given by
  \be\label{FRW}
  \left({d a \over d \eta}\right)^2= {c^2 a^2\over( H_0\eta)^2}+a\Omega^0_m + \Omega^0_r, 
  \ee
 which we will solve with the initial condition 
 \be
 a(\eta_0)=1\;
 \ee
to unfold the cosmology. In the above equations we have re-scaled the co-moving time by today's Hubble constant $H_0$, and the conformal time $\eta\equiv \int_t^\infty dt'/a(t')$ is different from the conventional one. $\Omega^0_m$ and $\Omega^0_r$ are the ratio of energy density of non-relativistic and relativistic matters to the critical density today.  Their values are $0.27$ and $8.23727\times 10^{-5}$, respectively, according to \cite{nova} and \cite{wmap}.  Note that relativistic matters here include both photon and neutrino. The same equation can also be expressed in the following form:
 \be
  {d \Omega_{\Lambda}(x) \over d x}= \Omega_{\Lambda}(x) \left\{1-\Omega_{\Lambda}(x)\right\} \left\{1+{2 \sqrt{\Omega_{\Lambda}(x)} \over c} + {\rho^0_r \over \rho^0_m {\rm e}^x +\rho^0_r} \right\}. 
  \ee
Here, $\rho^0_m$ and $\rho^0_r$ are the energy density of non-relativistic and relativistic matters today, and $x= {\rm ln}\, a.$  If we neglect $\rho^0_r$ (or $\rho^0_m$), then the equation can be solved  as in \cite{li1}. Otherwise one generally needs to solve the equation numerically. Since we are interested in the angular acoustic scale for which the radiation dominated epoch is also relevant, we must keep $\Omega_r$ and evolve the equation (\ref{FRW}) numerically. 
  
  To put constraints on HCDM, we first evaluate the acoustic oscillation in the CMB power spectrum of anisotropies. It occurred in the tightly coupled photon-baryon plasma before the recombination epoch, and then modulated the anisotropy of CMB spectrum observed today. The scale is characterized by \cite{hu} 
\be
\ell_A=\pi {d_* \over h_s}=\pi {\eta_0-\eta_{dec} \over \int_{\eta_{bb}}^{\eta_{dec}} c_s d\eta}.
\ee
Here, $d_*$ denotes the co-moving distance to the last scattering surface, $h_s$ the sound horizon right before the decoupling epoch, $c_s$ the sound speed, and $\eta_{bb}$ the co-moving time at the big-bang. Note that both $d_*$ and $h_s$ are affected by the presence of dark energy, as well as the matter and baryon density. The sound speed is given by
\be
c_s={1\over \sqrt{3(1+R)}},
\ee
with
\be
R\equiv {3\rho_b \over 4\rho_{\gamma}}\approx 30366\left(T^0_{\gamma}\over 2.725 K\right)^{-4} {\Omega^0_b h^2 \over 1+z}
\ee
where we have set $H_0=100h$ km s$^{-1}$ Mpc$^{-1}$. Apparently, the baryon-photon energy density ratio $R$ determines the ``stiffness constant'' in the acoustic oscillation of CMB.  

  On the other hand, the acoustic scale $\ell_A$ can be extracted by the locations of the acoustic peaks, which can be read out from the CMB power spectrum of anisotropy either theoretically \cite{mukhanov} or by a empirical formula \cite{doran}
\be
\ell_m=\ell_A (m-\varphi_m)
\ee
where $\ell_m$ denotes the angular position of the $m$-th acoustic peak. As explicitly shown in \cite{mukhanov} the acoustic oscillations for large $\ell$ are dressed by the dragging of the baryon-photon plasma and the Doppler shift due to the primordial gravitational bumps. This is why the empirical formula is a more direct way to extract the acoustic scale \cite{doran}. Using simulations, the authors have shown that the shift of the 3rd peak is relatively insensitive to the cosmological parameters and the best fit to its phase shift $\varphi_3$ is $0.341$. From the fitting of the BOOMERANG data at the year 2002 \cite{boom} they find that  
\be
\ell^{BOOM}_A=316\pm 8
\ee
with the following cosmological parameters: $h=0.65$ and $z_{dec}=1100$, the  matter density $\Omega^0_m=0.3$ which includes the baryon density $\Omega^0_b=0.05$, and $\Omega^0_{\Lambda}=0.7$. We would like to emphasize that this value of $\ell^{BOOM}_A$ has been extracted from a model-insensitive simulation \cite{doran}.

%fig1
\begin{figure}[ht]
\leavevmode
\hbox{
\epsfxsize=5in
\epsffile{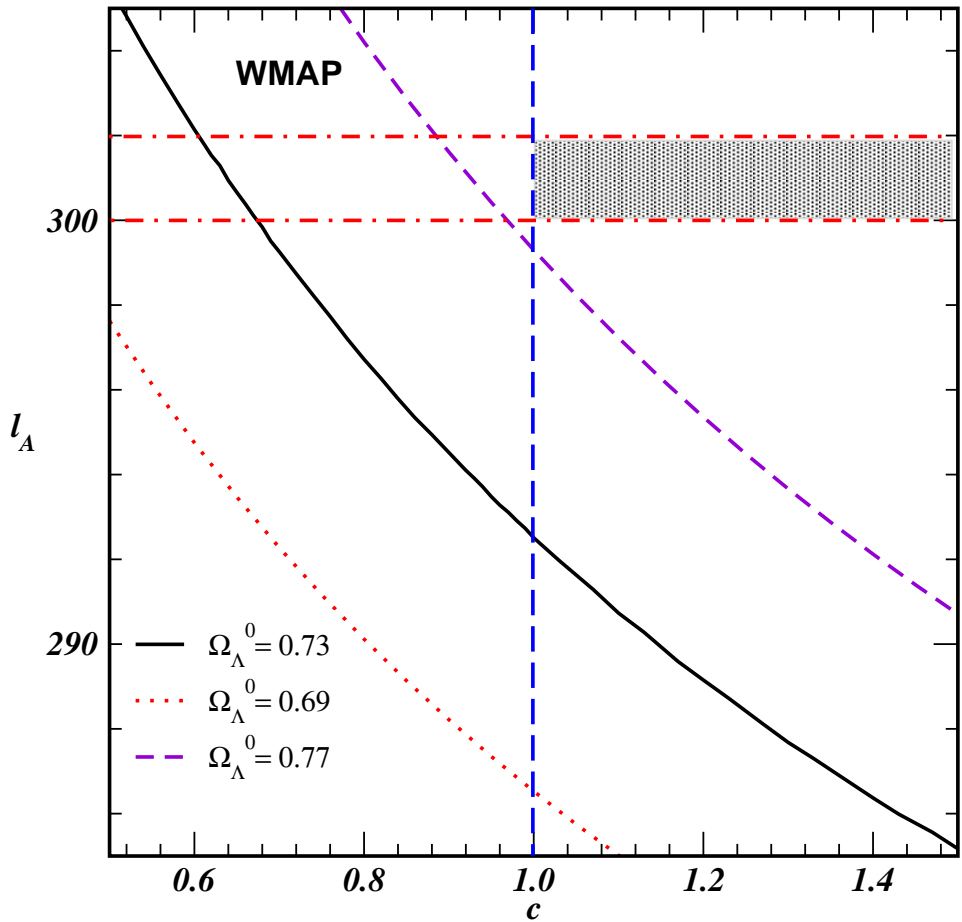}}
\caption{The solid curve depicts the dependence of $\ell_A$ on $c$ with $\Omega^0_r$ and $\Omega^0_m$ given by WMAP data. The shaded region is the 1-$\sigma$ window allowed by both the PEC/GSL and WMAP data. The dotted and the dashed curves below and above the solid one are obtained by taking into account the uncertainty of $\Omega^0_m$ while keeping $h$ and $\Omega^0_b$ fixed, as indicated in the graph.}
%\label{bg1}
\end{figure}

On the other hand, the 1st year WMAP data yield a value \cite{wmap}
\be
\ell^{WMAP}_A=301\pm 1
\ee
by fitting the underlying $\Lambda$CDM model with the following cosmological parameters: $h=0.71$ and $z_{dec}=1089$, $\Omega^0_m=0.27$ which include the baryon density $\Omega^0_b=0.044$, as well as $\Omega^0_{\Lambda}=0.73$.  Although the WMAP data is more up to date, no model-insensitive extraction of the acoustic scale $\ell_A$ has been done. To have a double check on HCDM, we will use both data sets to constrain HCDM.

  Using the cosmological parameters extracted from the WMAP data, we calculate the $\ell_A$ for HCDM models, and the result is presented in Fig. 1. The result shows that $\ell_A=292\pm7$ for $c=1$ which falls out of the window (the gray region) allowed by the WMAP data based on the underlying $\Lambda$CDM model and the PEC/GSL for the holographic entropy. 
%Moreover, those models residing within the allowed window but violating the GSL have $0.6 \lesssim c \lesssim 0.67$ which are phantom-like since $c^2<\Omega^{0}_{\Lambda}=0.73$ and thus $w^{0}_{\Lambda}<-1$. 
Since the allowed values for $\ell_A^{WMAP}$ is based on the underlying $\Lambda$CDM model, it cannot serve as a model-independent constraint. Our results derived from the WMAP data simply indicate how much the prediction by HCDM deviates from that by $\Lambda$CDM. Apparently, the deviation is quite significant for the non-phantom HCDM.

  Similarly the outcomes from the constraints by the BOOMERANG are displayed in Fig. 2.
  
%
%fig2
\begin{figure}[ht]
\leavevmode
\hbox{
\epsfxsize=5in
\epsffile{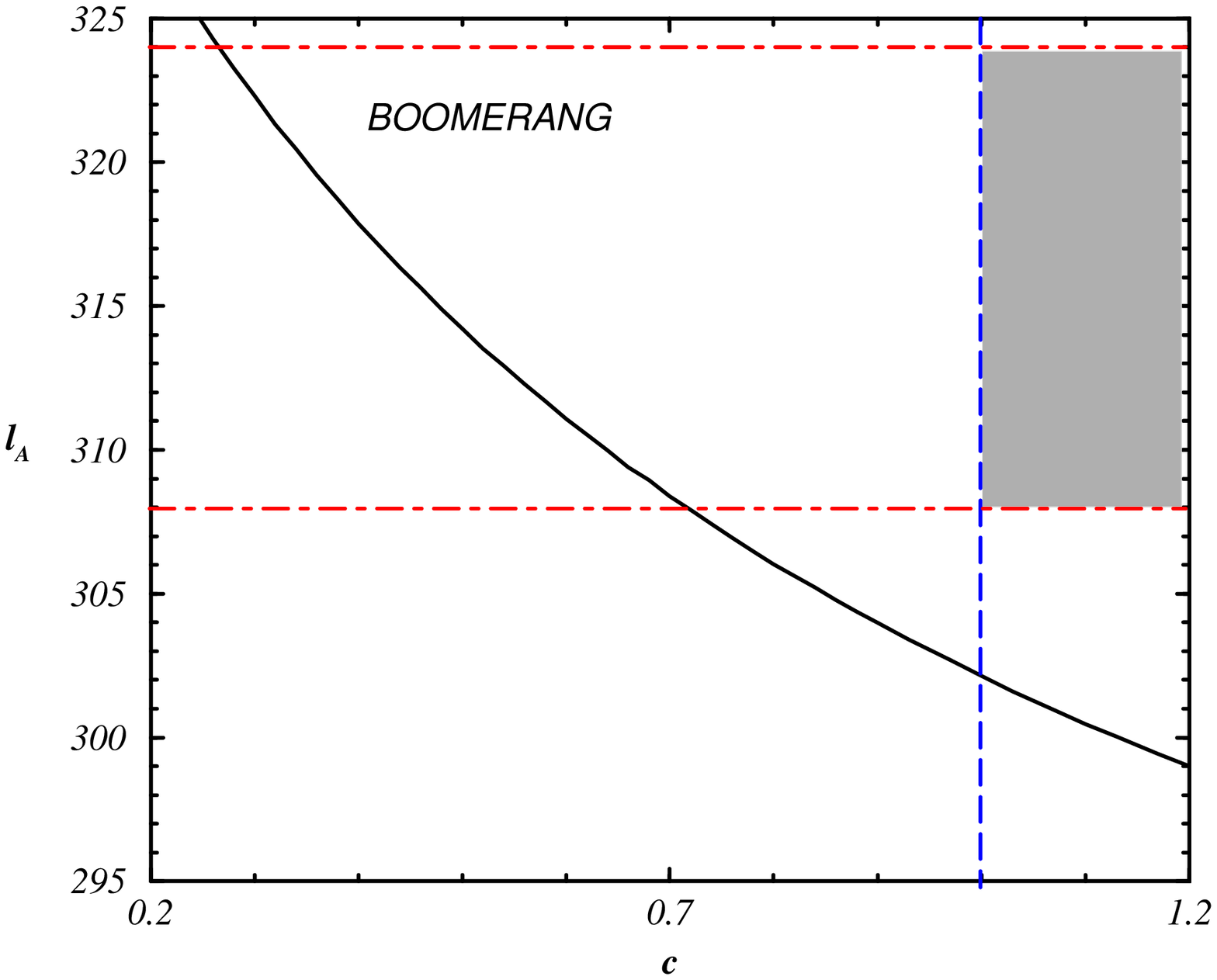}}
\caption{The solid curve depicts the dependence of $\ell_A$ on $c$ with $\Omega^0_r$ and $\Omega^0_m$ given by Boomerang data. The shaded region is the 1-$\sigma$ window allowed by both the PEC/GSL and Boomerang data. Unlike the WMAP, there is no estimate of error for the cosmological parameters in BOOMERANG.}
%\label{bg3}
\end{figure}
The results shows that $\ell_A=302$ for $c=1$ which still falls out of the window of BOOMERANG and PEC/GSL (the gray region) even though the window here is wider than the previous one from the WMAP data. 
%The values of $c$ within the BOOMERANG window but violating GSL ranges from $0.27$ to $0.71$ which are again phantom-like for $c^2<\Omega^{0}_{\Lambda}=0.7$. 
Since $\ell_A^{BOOM}$ for the BOOMERANG data is extracted from a model-insensitive method, it can serves as a model-independent constraint for HCDM. In summary, the above analyses of the acoustic angular scale from WMAP and BOOMERANG lead to the consistent conclusion that the HCDM models are disfavored.

%fig3
\begin{figure}[h]
\leavevmode
\hbox{
\epsfxsize=5in
\epsffile{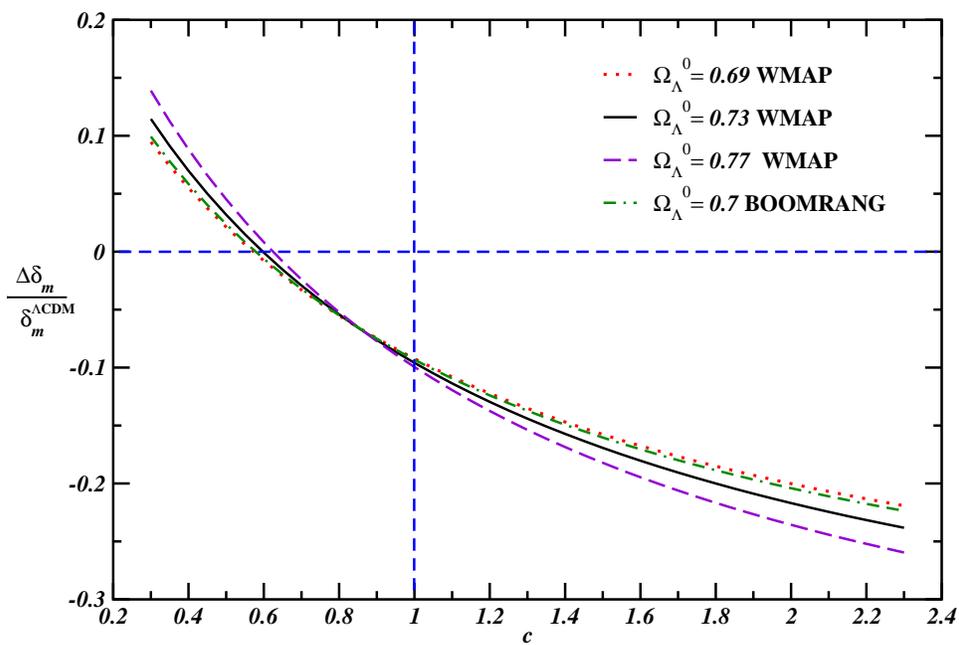}}
\caption{This figure shows the matter density suppression for different values of $c$. We denote the matter density parameters $\delta^{\rm HCDM}_m$, $\delta_m^{\Lambda{\rm CDM}}$ for HCDM and $\Lambda$CDM respectively, and $\Delta \delta_m \equiv \delta^{\rm HCDM}_m -\delta_m^{\Lambda{\rm CDM}}$. Different adoptions of the cosmological parameters are listed. Recall that a $30\%$ deviation  of $\delta^{\rm HCDM}_m$ from $\delta_m^{\Lambda{\rm CDM}}$ is allowed by the cosmic variance.}
%\label{bg2}
\end{figure}

  What we would like to do next is to check the constraint derived from the integrated Sach-Wolfe effect, which arises from the time-varying Newtonian potential (thus $w_{\Lambda}\neq 0$) after the last scattering surface. The ISW effect basically measures the breakdown of matter domination at early times (in the radiation regime) and at late times (in the dark energy regime).  The early ISW effect cannot be ignored if there exists a sizable amount of dark energy at the time of recombination. On the other hand, the late ISW effect becomes significant if the dark energy prevails along the way of evolution after recombination. In HCDM scenario, the dark energy density at decoupling time is negligible ($\sim 10^{-4}$) so that there is no early ISW effect. However, the late ISW effect for HCDM is substantial due to the running of dark energy. We will calculate the growth of the matter density contrasts $\delta$ for HCDM and $\Lambda$CDM to obtain the suppression of matter density in HCDM by comparing the two. A similar test has been performed for the generic quintessence scenario where the early ISW is significant (see \cite{leeng}). 
  
  Note that a more accurate determination would require computing small-scale matter perturbations as well as performing a maximum-likelihood fitting to the existing CMB anisotropy data. However, the constraint derived from the evolution of the growth factor under the influence of the ISW effect is more severe than the constraint from SN Ia data, and is good enough to serve as a fast tool to test any specific model of dark energy such as HCDM here. 
  
   The suppression effect actually is governed by the equation of motion 
\be
{d^2 \delta \over d^2 \eta} + a{\cal H} {d \delta \over d\eta} -{3\over 2} a^2 {\cal H}^2 \Omega_m \delta=0, 
\ee
where ${\cal H}\equiv H/H_0$ would inevitably admit the influence of the dark energy. The results are shown in Fig. 3. We find that the suppression increases with $c$. However, it is hard to use our results to constrain the HCDMs since the cosmic variance for low $\ell$ CMB spectrum is by itself large. If we take the cosmic variance as $30\%$ of the CMB normalization for the $\Lambda$CDM, then HCDM models with $c>1$ are allowed up to $c\sim 2.3$. In contrast to the acoustic peak scale, the ISW effect yields a loose bound on HCDMs due to the large cosmic variance. A peculiar thing in the result is the enhancement of the matter density $\delta^{\rm HCDM}_m$ for $c\lesssim 0.62$, i.e. $\Delta \delta_m \equiv \delta^{\rm HCDM}_m -\delta_m^{\Lambda{\rm CDM}}<0$. This may be due to the phantom nature of the dark energy. As emphasized previously, however, the phantom is in contradiction with the holographic principle.
  
%lin
   Our analysis shows that there is a discrepancy between the observational data of CMB and the theoretical preference of HCDM.  Though a phantom-like dark energy is by its own a viable model, it is inconsistent with the implicit PEC assumption in deriving HCDM and should be ruled out from the point of view of holographic principle. We thus conclude that the HCDM based on the UV-IR relations (\ref{bound1}) and ({\ref{rhoL}) is marginally ruled out by present CMB data. As known, the first year WMAP data precisely measure the power spectrum to $\ell=300$, which is marginally good in determining the acoustic peak sclae. One may expect the more updated CMB data to clear the mess, for example, the coming WMAP data is expected to reach $\ell=600$ \cite{spe}.

%lin   
   More specifically, the data show that the model provides too much dark energy than necessary, one may then wonder if we can reduce the amount of holographic dark energy by setting an onset time. Physically, this can be understood from the possibility that the dark energy is the latent heat dumped during the cosmic phase transition such as the electroweak or QCD ones, and it takes effect in cosmic evolution only after some onset times $z_{\ast}$. This provides one more parameter bedsides $c$. However, by repeating similar analysis we find that $z_{\ast}$ must be greater than the decoupling time to make the $c>1$ HCDM survive the CMB constraint on acoustic scale. This is counter-empirical since we know none of the fundamental phase transition with energy scale smaller than the atomic scale. Therefore, we conclude that HCDM is not viable even when we introduce an onset time for it.
   
    Since the argument for holographic principle is quite general, and we expect that it plays a crucial role in defining the dark energy. Hence, the inconsistency shown in this paper indicates that there may be some subtlety in implementing MDC. Knowing that our universe is not as dark as inside a black hole, we have abandoned the old covariant entropy bound and adopted the bound on Jeans instability, which lead us to the HCDM model.  It is quite possible that our universe is not even dark enough to saturate Jeans' bound, and we may need a new criterion to relate the UV and IR physics in implementing MDC based on the holographic principle.

   \bigskip

{\bf Acknowledgment}: FLL would like to thank the warm hospitality of Chuo University and KEK where part of manuscript was drafted during his visit, and also Ko Furuta, Takeo Inami, Miao Li for discussions. WLL thanks Kin-Wang Ng and Eric Linder for helpful discussions.

\end{document}